\documentstyle[12pt]{article}
\begin{document}
\begin{center} \huge {TOP NEAR THRESHOLD:}
\end{center}
\begin{center}
      \huge {all ${\alpha}_{S}$ corrections are trivial.}
\end{center}
\vspace{0.2cm}
\begin{center} \bf {K. Melnikov$^{1,3}$ and O. Yakovlev$^{2,3}$}
\end{center}
\vspace{0.1cm}
\begin{center} $^{1}$ Novosibirsk State University , 630090
Russia, Novosibirsk \\
 $^{2}$ Budker Institute for Nuclear Physics , 630090
Russia, Novosibirsk \\ e-mail: O$\_$Yakovlev@INP.NSK.SU \\
 $^{3}$ Supported in part by grant from Soros fund "Culture initiative"\\
          preprint - BUDKERINP 93-18
\end{center}
\vspace{0.2cm}
\begin{center} \bf {Abstract}
\end{center}
    We have calculated part of  O(${\alpha}_{S}$) QCD radiative
    correction to the total cross-section of top's threshold
    production in $ e^{+}e^{-} \to t\bar t \to W^{-} b W^{+} \bar b$
    reaction. We found out a curious fact: there is no  O(${\alpha}_{S}$)
    correction to the total cross-section, originated from $b \bar b ,
    t \bar b ,\bar t b $ gluon exchange and corresponding gluon emission.
    Thus, all  O(${\alpha}_{S}$) corrections which are non-zero, can be
    classified as: 1) corrections to $t \bar t$ nonrelativistic potential
    2) corrections to the top's width
    3) hard gluon correction to $\gamma t \bar t$-vertex.
    These corrections are well-known and were described in the
    previous papers.
     We can conclude that similar situation takes
    place for all unstable particles threshold production. This result
    can be expressed in the form of a general statment:
\begin{center}{ There is no O(${\alpha}_{QED}$  or ${\alpha}_{QCD}$)
corrections for the total cross-section , originated from the
"jet-jet" interection.}
\end{center}

\begin{center} {\bf 1. Introduction}
\end{center}
\par
 The problem of top production near threshold has been posed and
    solved by Fadin and Khoze [1].This and following papers [2-6]
    led people to conclusion that it would be possible to measure
    both top's mass and width and strong coupling constant ${\alpha}_S$
    using top's threshold production. This belief is based on some
    important grounds:
    \begin{itemize}
    \item non-perturbative corrections are small [1,2,3]
    \item runnig of QCD coupling constant can be consistently
    taking into account [3]
    \item effects originated from top's width dependence from
    virtuality are in principle calculacable [4]
    \end{itemize}
    On this way an important problem has not been solved yet - it is
    the problem of O(${\alpha}_{S}$) QCD radiative corrections
    calculation. In this letter we are going to study this problem
    in details.
\par
    Let us firstly remind the reader about results obtained in the
    leading order approximation in ref.[1]. Summing "Coulomb-like"
    ladder Fadin and Khoze obtained the following cross-section for
    top's threshold production
    \begin{eqnarray}
    \sigma (e^{+}e^{-} \to t\bar t \to W^{-} b W^{+} \bar b) =
    N_{c}{Q_{t}}^{2}\sigma (e^{+}e^{-} \to \mu \bar \mu)
    \frac{4\pi}{s} Im{P}_{t}(s) \nonumber \\
    \sigma (e^{+}e^{-} \to \mu \bar \mu )=
    \frac{4\pi{\alpha}^{2}_{QED}}{3s}
    \end{eqnarray}
    Here $Q_{t}$is the top's charge and ${P}_{t}(s)$ stands for
    vacuum polarization by top's vector current.\\ $s=q^{2}$,
    ${P}_{t}^{\mu \nu}(s)={P}_{t}(s)(\frac{q^{\mu}q^{\nu}}
    {q^{2}}-g^{\mu \nu})$.
    ${P}_{t}(s)$ for
    $\mid s-4m_{t}^{2}\mid \ll 4m_{t}^{2} $ is connected with the
    non-relativistic Green function of Coulomb-like problem:
    \begin{eqnarray}
    {P}_{t}(s)=\frac{3}{2{m_{t}}^{2}} G_{E+i\Gamma}(0,0) \\
    G_{E+i\Gamma}(r,0)=<\vec r\mid \frac{1}{\hat H -E-i\Gamma}\mid \vec r=0> =
    \int \frac{d\vec p}{{(2\pi)}^{3}}
    \frac{V(E,\vec p){e}^{i\vec p \vec r}}
    {E-\frac{{\vec p}^{2}}{m}+i\Gamma} \\
    \hat H = \frac{{\vec p}^{2}}{m}-\frac{4}{3}
    \frac{{\alpha}_{S}}{r} \nonumber \\
    \end{eqnarray}
    Here $E=\sqrt{s}-2m_{t}$ and $V(E,\vec p)$ is the
    $\gamma t \bar t$ nonrelativistic vertex function (Fig.1). \\
    We'll use all this notations throughout the paper. \\
    What we are going to do is to study O(${\alpha}_{S}$) QCD radiative
    corrections to the Eq.(1). On this way, one must recognize that
    there is no additional "smallness" due to electroweak coupling
    constants in the leading order cross-section because they are
    compensated by top's resonanse propagators. Hence, we are interested
    in a correction, that does not shift quark's propogator out of
    the pole. Dealing with "jet-jet" interaction, we see that this
    requirement results in the fact that we must consider "soft"
    gluon contribution ; the required "softness", however, depends
    greatly from the diagram in question.

\begin{center}{\bf 2. Method of calculation }
\end{center}
\par
    As we intend to study the correction to the total
    cross-section, it is convinient to choose the the following way of
    work (for concreteness let us consider the typical graph, depictured
    in Fig.2): firstly, integrate over the phase-space of $b W^{+}$
    ("cut" block A) and $\bar b W^{-}$ ("cut" block B).
    On this way one'll obtain two quanaties: $I_{\mu}$ for the upper
    block and $J_{\mu}$ for the lower one (Fig.2a).
    Following ideas of ref.[1], the non-relativistic approximation can
     be used to simplify this quantaties. Secondly, write down the
     amplitude of Fig.2a in term of  $I_{\mu}$ and $J_{\mu}$ and
     integrate over loop momentum and top's virtuality.
     \par
     As we are not far from threshold, it is sufficient to use
 nonrelativistic approximation [1] for $t$ and $\bar t$ propogators:
\begin{equation}
S_{t}(p_{t})=\frac{1+{\gamma}_{0}}{2} \frac{1}{{\varepsilon}_{t}-
\frac{{\vec p}^{2}}{2m_{t}}+i\frac{{\Gamma}_{t}}{2}}
\end{equation}
\begin{equation}
S_{\bar t}(-p_{\bar t})=\frac{1-{\gamma}_{0}}{2}
\frac{-1}{{\varepsilon}_{\bar t}-
\frac{{\vec p}^{2}}{2m_{t}}+i\frac{{\Gamma}_{t}}{2}}
\end{equation}
Here   $ p_{t}=(m_{t}+{\varepsilon}_{t}, \vec p ) $  ,
$ p_{\bar t}=(m_{t}+{\varepsilon}_{\bar t},- \vec p ) $ is top's
four momentum.
We ignore in (5),(6) the momentum-dependence of the top quark width.
\par
Let us now turn to the discussion of different diagrams contribution
to the radiative correction. Studing various types of "jet-jet"
 interaction we found it convinient to use the Coulomb gauge. We also
found the same result in the Feynman gauge.
\begin{center} {\bf 3. Contribution of different diagrams to the
         total cross-section.}
\end{center}
\begin{center} {\bf 3.1 Subgraphs calculation.}
\end{center}
First of all, let us culculate $I_{\mu}$ and $J_{\mu}$ - quantaties,
introduced earlier. Applying Feynman rules, we get in nonrelativistic
 approximation:\\
1) for time-like gluon:
\begin{eqnarray}
{I}_{0} & = & -\sqrt{4\pi{\alpha}_{S}}\hat w_{+}
\frac{{\Gamma}_{t}}{k} \nonumber\\
& & \Biggl\{\frac{1}{2}ln\Biggl(\frac{w-k+i\varepsilon}
{w+k+i\varepsilon}\Biggr)
\gamma_{0}-
\frac{2x-1}{2x+1}\Biggl(1+\frac{w}{2k}
ln\Biggl(\frac{w-k+i\varepsilon}{w+k+i\varepsilon}
\Biggr)
\frac{\vec \gamma \vec k}{k} \Biggr\}
\end{eqnarray}

2) for space-like gluon :
\begin{equation}
{I}_{i} =  \sqrt{4\pi{\alpha}_{S}}\hat w_{+}
\frac{{\Gamma}_{t}}{2k} \gamma_{i}
\frac{(2x-1)}{(2x+1)}
\Biggl\{\frac{w^2-k^2}{2k^{2}}
ln\Biggl(\frac{w-k+i\varepsilon}{w+k+i\varepsilon}
\Biggr)
+{\frac{w}{k}}
\Biggr\}
\end{equation}
Here $ k= \mid \vec k \mid  ; w ,
\vec k $ -energy and momentum of gluon,
$ x={\frac{m_{W}^2}{m_{t}^{2}}} $ ,
${\Gamma}_{t}=m_{t}{g_{W}}^{2}\frac{(1-x)^{2}(1+2x)}{8\pi x}$ -top's width,
${\alpha}_{S}$ - strong coupling constant and
$\hat w_{+} =1+{\gamma}_{5} $. \\
For further calculations it is convinient to introduce the following
notations:
\begin{equation}
{I}_{0}=w_{+}{\gamma}_{0}A+w_{+}{\gamma}_{i}B^{i},\qquad
{I}_{i}=w_{+}{\gamma}_{i}C
\end{equation}
Straightforward calculation give
\begin{equation}
{J}_{0}=w_{+}{\gamma}_{0}{A}^{*}+w_{+}{\gamma}_{i}{B^{i}}^{*},
\qquad
{J}_{i}=w_{+}{\gamma}_{i}{C}^{*}
\end{equation}
There is a set of "cut" subgraph
$ \bar I(p_{t},k) $ and $ \bar J(p_{t},k) $ of similar nature, presented
at Fig.2b (Block D,C). They are connected with
$I(p_{t},k) $ and $J(p_{t},k) $ in following way:
\begin{equation}
{\bar J}_{\mu}=I_{\mu} \qquad  {\bar I}_{\mu}={ J}_{\mu}
\end{equation}
\newpage
\begin{center} {\bf 3.2 Time-like gluon exchange
 between $ b \bar b $ - quarks}
\end{center}
\par
In this case, we deal with the graphs in Figs.2a,b. Their contribution
to the function $Im P_{t}$ reads:
$Im P_{t}^{b \bar b}=\frac{2}{3}Re M^{b \bar b}(E).
( Here  M^{b \bar b}=T^{*(one-loop)}T^{(Born)}$ and $T$
stands for the
amplitude of the process
${\gamma}^{*}\to t\bar t \to W^{-} b W^{+} \bar b)$.
 Here $M^{b \bar b}$ reads:
\begin{eqnarray}
{M}^{b \bar b} = -i\int \frac{d^{4}p}{{(2\pi)}^{4}}
\int \frac{d^{4}k}{{(2\pi)}^{4}}
C_{F}
{\bf Sp}\Biggl({\gamma}_{i}V^{*}(\vec p){S^{*}}_{t}(p_{t})I_{0}
\nonumber \\
S_{t}(p_{t}+k){\gamma}^{i}V(\vec p+\vec k)
S_{\bar t}(-p_{\bar t}+k)J_{0}
{S^{*}}_{\bar t}(-p_{\bar t})\Biggr)
D(\vec k)
\end{eqnarray}
where $ C_{F}=\frac{N_{c}^{2}-1}{2N_{c}},
D(\vec k)= \frac{1}{{\mid \vec k \mid}^{2}} $
stands for photon propogator. Integration over
 $ d\varepsilon $ can be performed explicitly by taking two residues
 in the points:
\begin{equation}
\varepsilon=\frac{{\vec p}^{2}}{2m_{t}}+i\frac{{\Gamma}_{t}}{2},
\qquad
\varepsilon=
E-w-\frac{{(\vec p+\vec k)}^{2}}{2m_{t}}+i\frac{{\Gamma}_{t}}{2}
\end{equation}
\par
After that  $ {M}^{b \bar b} $ takes the form:
\begin{eqnarray}
M^{b \bar b} = -\int \frac{d^{3}p}{{(2\pi)}^{3}}
\int \frac{d^{3}k}{{(2\pi)}^{3}}
\int \frac{dw}{{(2\pi)}}
\Biggl(3A{A}^{*}+\vec B {\vec B}^{*}\Biggr)
\frac{C_{F}}{{\mid \vec k \mid}^{2}}
 \frac{4\Delta}{w^{2}-{\Delta}^{2}} \nonumber\\
\frac{V^{*}(\vec p)}{E-\frac{{\vec p}^{2}}{m_{t}}-i{\Gamma}_{t}}
\frac{V(\vec p +\vec k)}
{E-\frac{{(\vec p+\vec k)}^{2}}{m_{t}}+i{\Gamma}_{t}}
\end{eqnarray}
Here $ \Delta = - \frac{{(\vec p+\vec k)}^{2}}
{2m_{t}}+
\frac{{(\vec p)}^{2}}
{2m_{t}}+i{\Gamma}_{t}$ and $A, \vec B , C$  are introduced
through Eqs.(7-9). \\
Simple dimensional analyses leads us to conclusion, that the only
integration region,
 wich can give correction of O(${\alpha}_{S}$) order
is the region $k, w < \Gamma$.\\
For example, the region $w \sim \Gamma  ,  k \sim p_{t} \sim
{\alpha}_{S}m_{t}$ will give  ${\alpha}_{S}
\frac{{\Gamma}_{t}}{p_{t}} \sim \frac{{\Gamma}_{t}}{m_{t}}
\sim {\alpha}_{W} $,
 where ${\alpha}_{S}$ is the electroweak coupling
 constant (${\alpha}_{W}\sim {\alpha}_{S}^{2}$).\\
 Hence, we are only interested in the integration over region
 $k, w < \Gamma$ , but it is not difficult to see from Eq.[14]
 that in this case $M^{b \bar b}$ is purely imaginary
( $ \Delta =i{\Gamma}_{t} $). Hence
 $Im P_{t}^{b \bar b}=\frac{2}{3}Re M^{b \bar b}$ appears to be zero up to
 O(${\alpha}_{S}$) order.

\begin{center} {\bf 3.3. Time-like gluon exchange between $t$ and $\bar b$
 ($\bar t$ and $b$) -quarks.}
\end{center}
\par
 Different diagrams corresponding to this case are presented in Figs.
 3a,b.
Their contribution to the total cross-section appears to be identical.
This fact can be checked by straightforward calculation or by applying
charge conjugation to one of these diagrams.
The contribution to $Im P_{t}(E)$ reads:
\begin{eqnarray}
Im P_{t}^{t \bar b}  =  \frac{4}{3}Re M^{t \bar b}(E) \nonumber\\
{M}^{t \bar b}  =  -i\int \frac{d^{4}p}{{(2\pi)}^{4}}
\int \frac{d^{4}k}{{(2\pi)}^{4}}
C_{F}
{\bf Sp}\Biggl({\gamma}_{i}V^{*}(\vec p){S^{*}}_{t}(p_{t})\hat {\Gamma}
{S}_{t}(p_{t}) \nonumber \\
(\sqrt{4\pi {\alpha}_{S}}{\gamma}_{0})
S_{t}(p_{t}+k){\gamma}^{i}V(\vec p+\vec k)
S_{\bar t}(-p_{\bar t}+k)J_{0}{S^{*}}_{\bar t}(-p_{\bar t})\Biggr)
D(\vec k)
\end{eqnarray}
Here $\hat {\Gamma}=w^{+}{\gamma}_{0}{\Gamma}_{t} $.
 Performing the trace and
integrating over $d\varepsilon$, we get
\begin{eqnarray}
{M}^{t \bar b} = - \int \frac{d^{3}p}{{(2\pi)}^{3}}
\int \frac{d^{3}k}{{(2\pi)}^{3}}
\int \frac{dw}{{(2\pi)}}
\frac{(\sqrt{4\pi {\alpha}_{S}}A^{*})}{k^{2}}
\frac{12\Delta}{w^{2}-{\Delta}^{2}} \nonumber\\
\frac{V^{*}(\vec p_{t})}
{E-\frac{{\vec p}^{2}}{m_{t}}-i{\Gamma}_{t}}
\frac{V(\vec {p_{t}}+\vec k)}
{E-\frac{{(\vec p+\vec k)}^{2}}{m_{t}}+i{\Gamma}_{t}}
\end{eqnarray}
In Eq.(16), we omit the terms which after integration over $w$ are zero.
We can perform integration over $w$ taking into account analytical
properties of the logarithm
$ln\Biggl(\frac{w-k+i\varepsilon}{w+k+i\varepsilon}\Biggr)$.
After that one obtain:
\begin{equation}
{M}^{t \bar b}=i
 \int \frac{d^{3}p}{{(2\pi)}^{3}}
\int \frac{d^{3}k}{{(2\pi)}^{3}}
\Phi(\mid\vec k\mid)
G^{*}(\vec r=0,\vec p \mid E+i{\Gamma}_{t})
G(\vec r=0,\vec p+\vec k\mid E+i{\Gamma}_{t})
\end{equation}
The function $\Phi (k)$ reads here:
\begin{equation}
\Phi(\mid\vec k\mid)=
12\pi{\alpha}_{S}\frac{{\Gamma}_{t}}{k^{3}}(\pi-2arctg(\frac{{\Gamma}_{t}}{k})
)
\end{equation}
 As is seen from Eq.(14) the main contribution come from the
 region $k \sim p \sim m_{t}{\alpha}_{S}$;
 thus we can't neglect $k$ in comparison with $p$ in
 the argument of Green function. However, we can overcome this
 difficulty
 inserting full set of state ($\mid \vec r><\vec r\mid=1$)
and after that:
\begin{equation}
{M}^{t \bar b}=i
\int {d^{3}r}
{\mid G(\vec r=0,\vec r\mid E+i\Gamma)\mid}^{2}
\int \frac{d^{3}k}{{(2\pi)}^{3}}
\Phi(\mid\vec k\mid) e^{i\vec k\vec r}
\end{equation}
So, we see, that ${M}^{t \bar b}$ appears to be pure imaginary, hence
these diagrams give no contribution to O(${\alpha}_{S}$) correction.

\begin{center} {\bf 3.4 Space-like gluon exchange between $b \bar b$
 and corresponding emission of real gluon.}
\end{center}
 Let us now study the space-like gluon exchange. The first note is
 that we can attach space-like gluon only to $b$-quark, because,
 as is well-known, space-like gluon interection to quark current
 is proportional to quarks velocity which (for the top) is of
 O(${\alpha}_{S}$) order, while for $b$ quark it is of order 1.
 That's why we consider the graphs with space -like gluon exchange
 between $b \bar b$, and corresponding gluon emission Figs.2a,b,4a,b.
In latter
 case there are two possibilities:  1)  the interference between
 radiation from $b$ and $\bar b$   2)  $b $ ($\bar b$) radiation.
 Our interest
 is concentrated on the first possibility, because the second one
 is included to the correction to $t \to Wb$
 ($\bar t \to W\bar b$) width.
 The contribution to $Im P_{t}(E)$ reads:
\begin{eqnarray}
Im{P}^{b \bar b}_{sp}=\frac{2}{3}Re{M}^{b \bar b}_{sp}
+\frac{2}{3}Re{M}^{b \bar b}_{real}
\nonumber\\
{M}^{b \bar b}_{sp}= - \int \frac{d^{4}p}{{(2\pi)}^{4}}
\int \frac{d^{4}k}{{(2\pi)}^{4}}
 \frac{1}{3}
{\bf Sp}\Biggl({\gamma}_{l}V^{*}(\vec p_{t}){S^{*}}_{t}(p_{t})I_{i}
\nonumber\\
S_{t}(p_{t}+k){\gamma}^{l}V(\vec p_{t}+\vec k)
S_{\bar t}(-p_{\bar t}+k)J_{j}{S^{*}}_{\bar t}(-p_{\bar t})\Biggr)
D^{i j}(\vec k,w) \\
{M}_{real}^{b \bar b} = - \int \frac{d^{4}p}{{(2\pi)}^{4}}
\int \frac{d^{3}k}{{(2\pi)}^{3}2w}
C_{F}
{\bf Sp}\Biggl({\gamma}_{l}V^{*}(\vec p_{t}){S^{*}}_{t}(p_{t})I_{i}
\nonumber \\
S_{t}(p_{t}+k){\gamma}^{l}V(\vec p_{t}+\vec k)
S_{\bar t}(-p_{\bar t}+k)\bar J_{j}{S^{*}}_{\bar t}(-p_{\bar t})\Biggr)
e^{* i} e^{j}
\end{eqnarray}
${M}^{b \bar b}_{sp}$ is the virtual gluon contribution,
${M}^{b \bar b}_{real}$ is the radiation interference contribution.
Also $ D^{i j}(\vec k,w)= \frac{i}{ w^{2} - {\mid \vec k \mid}^{2}+i0}
({\delta}^{i j} - {n^{i}n^{j}}) $,
 $ \vec n = \frac{\vec k}{\mid\vec k\mid}$, $e^{\mu}=(0,\vec e)$ -
 wave function of gluon, for summation over
polarization of gluon, we use formula $ \Sigma e_{i}e^{*}_{j}=(\delta_{i j}
-n_{i}n_{j}) $.\\
Simple dimensional analyses leads us to conclusion, that the only
integration region,
 wich can give correction of O(${\alpha}_{S}$) order
is the region $k, w < \Gamma$.\\
 Performing the trace, and integrating over the energy of the top, we get:
\begin{eqnarray}
Im{P}^{b \bar b}_{sp}=-Re
\int \frac{d^{3}p}{{(2\pi)}^{3}}
\int \frac{d^{3}k}{{(2\pi)}^{3}}
\frac{16{\Gamma}_{t}}{3}\frac{V^{*}(\vec p_{t})}
{E-\frac{{\vec p}^{2}}{m_{t}}-i{\Gamma}_{t}}
\frac{V(\vec {p_{t}}+\vec k)}
{E-\frac{{(\vec p+\vec k)}^{2}}{m_{t}}+i{\Gamma}_{t}} \nonumber \\
 \Biggl\{
\int \frac{dw}{{(2\pi)}}
\frac{C(w,\vec k)C^{*}(w,\vec k)}{(w^{2}+{{\Gamma}_{t}}^{2})}
\frac{i}{ w^{2} - {\mid \vec k \mid}^{2}+i0}
 -  \frac{C(w=\mid\vec k\mid)C^{*}(w=\mid\vec k\mid)}
{2k(k^{2}+{{\Gamma}_{t}}^{2})}\Biggr\}
\end{eqnarray}
Integration over $w$ can be perfomed by the use of the well-known
formulae:
$\int dx \frac{1}{x+i0}=P(\frac{1}{x})-i\pi\delta(x)$.
The principal value does not contribute due to the fact that it is
purely imaginary, while the pole part of the virtual correction
cancel the real emission contribution.
\begin{center}{\bf 3.5 Other corrections.}
\end{center}
There are other corrections, originated from the space-like gluon
exchange between b-quarks and Coulomb gluon , and real gluon emission
from Coulomb gluon. Using techniquediscussed above, one can conclude
 that they don't give O(${\alpha}_{S}$) correction to the total
cross-section.
\begin{center} { \bf 4.Conclusion.}
\end{center}
\par
So we have considered the O(${\alpha}_{S}$) QCD radiative correction to
top threshold production cross-section
$e^{+}e^{-} \to t\bar t \to W^{-} b W^{+} \bar b$, originated from
"jet-jet" interection in the final state. We found out that thous
corrections give nothing in O(${\alpha}_{S}$) order.
Hence, O(${\alpha}_{S}$) radiative correction appears only from:
    \begin{enumerate}
    \item corrections to $t \bar t$ nonrelativistic potential.
    \item corrections to the top's width.
    \item hard gluon correction to $\gamma t \bar t$-vertex
    \end{enumerate}
Thus all they are trivial in the sence , that we can use  the
general formula for the total cross-section obtained in the leading
order and then substituting
${V(r)}^{coulomb} \to {V(r)}^{one-loop} ,
{\Gamma}^{Born} \to {\Gamma}^{one-loop} ,
\sigma^{Born}(e^{+}e^{-} \to \mu \bar \mu) \to
\sigma^{Born}(e^{+}e^{-} \to \mu \bar \mu)
(1+\frac{8{\alpha}_{S}}{3\pi}),$
we obtain the top's threshold production cross-section
up to O(${\alpha}_{S}$) order.
\begin{eqnarray}
\sigma (e^{+}e^{-} \to t\bar t \to W^{-} b W^{+} \bar b) =
\sigma^{Born}(e^{+}e^{-} \to \mu \bar \mu)
(1+\frac{8{\alpha}_{S}}{3\pi})
\frac{6\pi Q_{t}^{2}}{m_{t}^{2}} \nonumber\\
Im<r=0\mid \frac{1}{\hat H -E-i{\Gamma}^{one-loop}}\mid r=0> \\
\hat H = \frac{{\vec p}^{2}}{m}-{V(r)}^{one-loop} \nonumber
\end{eqnarray}
This is the general answer for the problem, outlined in the
introduction. \\
{}From above discussion , we can conclude that similar situation takes
 place for all unstable particles threshold production. This result
 can be expressed in the form of a general statment:
\begin{center}{ There is no O(${\alpha}_{QED}$  or ${\alpha}_{QCD}$)
corrections for the total cross-section , originated from the
"jet-jet" interection.}
\end{center}
We want to add in conclusion, that , for our mind, the absence of
O(${\alpha}_{S}$) correction to the total cross section, originated
 from "jet-jet" interection, must have simple physical meaning
and now we are looking for it.
\begin{center} { \bf Acknowledgement:}
\end{center}
 The authors are very gratefull to Profs. V.S.Fadin , I.F.Ginzburg
 and Dr. A.G.Grozin for discussions.

\end{document}